\documentclass[fp,twocolumn]{jpsj3}
\usepackage{txfonts}
\usepackage{graphicx}
\usepackage{dcolumn}
\usepackage{bm}
\usepackage[usenames,dvipsnames]{xcolor}
\usepackage{ulem}
\usepackage{amsmath}

\title{Spontaneous multipole ordering by local parity mixing}

\author{Satoru Hayami$^1$\thanks{E-mail address: hayami@aion.t.u-tokyo.ac.jp}, Hiroaki Kusunose$^2$, and Yukitoshi Motome$^1$}
\inst{$^1$Department of Applied Physics, University of Tokyo, Tokyo 113-8656, Japan \\
$^2$Department of Physics, Ehime University, Matsuyama 790-8577, Japan 
}

\abst{
Broken spatial inversion symmetry in spin-orbital coupled systems leads to 
a mixing between orbitals with different parity, which results in unusual electronic structures and transport properties. 
We theoretically investigate the possibility of multipole ordering induced by a parity mixing. 
In particular, we focus on the system in which the parity mixing appears in a sublattice-dependent form. 
Starting from the periodic Anderson model with such a local parity mixing, we derive an extended Kondo lattice model with sublattice-dependent antisymmetric exchange couplings between itinerant electrons and localized spins. 
By the variational calculation, simulated annealing, and Monte Carlo simulation, 
we show that the model on a quasi-one-dimensional zig-zag lattice exhibits an odd-parity multipole order composed of magnetic toroidal and quadrupole components at and near half filling. 
The multipole order causes a band deformation with the band bottom shift and a magnetoelectric response. 
The results suggest that unusual odd-parity multipole orders will be widely observed in multi-orbital systems with local parity mixing. 
}

\begin{document}
\maketitle

\section{Introduction}
\label{sec:Introduction}

The antisymmetric spin-orbit coupling, which is induced by inversion symmetry breaking of the crystal structure, has attracted interest because it leads to fascinating phenomena, such as the magnetoelectric effect~\cite{curie1894symetrie,dzyaloshinskii1960magneto,kimura2003magnetic,wang2003epitaxial,KhomskiiPhysics.2.20} and  noncentrosymmetric superconductivity~\cite{Bauer_PhysRevLett.92.027003,Bauer_Sigrist201201,Frigeri_PhysRevLett.92.097001,Samokhin_PhysRevB.69.094514,Fujimoto_PhysRevB.72.024515,Fujimoto_doi:10.1143/JPSJ.75.083704}. 
In such noncentrosymmetric systems, the Hamiltonian for the antisymmetric spin-orbit coupling is written by 
\begin{align}
\label{eq:Ham_ASOC_def}
\mathcal{H}_{{\rm ASOC}} (\bm{k})
=\bm{g}(\bm{k}) \cdot \bm{\sigma} \sim (\bm{k} \times \nabla V_{{\rm pot}}) \cdot \bm{\sigma}, 
\end{align}
where $\bm{g}(\bm{k})$ is the polar vector with respect to the wave vector $\bm{k}$, 
$\bm{\sigma}$ is the spin operator, and $\nabla V_{{\rm pot}}$ is a potential gradient due to the inversion symmetry breaking.
This antisymmetric spin-orbit coupling often provides a clue for understanding of peculiar electronic and transport properties, such as the spin splitting in the band structure~\cite{rashba1960properties,bychkov1984oscillatory,Dresselhaus_Dresselhaus_Jorio}. 
 
In such spin-orbital coupled phenomena, not only uniform but also local (site) inversion symmetry breaking has recently drawn considerable attention~\cite{sakhnenko2012magnetoelectric,ter2014cation,Yanase:JPSJ.83.014703}.
The local inversion symmetry breaking means that the inversion symmetry is broken at every lattice site in a different manner from site to site. 
For instance, a two-dimensional honeycomb lattice breaks the local inversion symmetry at every lattice site in an alternating way on the two sublattices, while it preserves the inversion symmetry at each center of the nearest-neighbor bonds and the hexagons. 
In these systems, the parity mixing between different orbitals occurs locally in a sublattice-dependent form, which we call the local parity mixing. 
In the presence of the local parity mixing, $\bm{g}(\bm{k})$ in Eq.~(\ref{eq:Ham_ASOC_def}), also depends on the sublattice. 
This suggests that new spin-orbital coupled phenomena are expected by using the sublattice degree of freedom in crystals. 
For example, once a sublattice-dependent magnetic or electronic order is induced in the system with local parity mixing, multipoles extending over different sublattice sites are simultaneously activated in odd-parity sectors, such as a magnetic quadrupole and an electric octupole~\cite{Spaldin:0953-8984-20-43-434203,kopaev2009toroidal}. 

Such unusual odd-parity multipoles have been recently studied for unconventional superconductivity~\cite{Yoshida_doi:10.7566/JPSJ.82.074714,Yoshida_PhysRevB.86.134514,Maruyama_doi:10.1143/JPSJ.81.034702} and electronic orders~\cite{Yanase:JPSJ.83.014703,Hayami_PhysRevB.90.024432,Hayami_PhysRevB.90.081115,hayami2014quantum,hayami2014toroidal,hitomi2014electric}. 
A typical case was theoretically discussed on a zig-zag chain, where the local inversion symmetry is broken and the sublattice-dependent antisymmetric spin-orbit coupling is present~\cite{Yanase:JPSJ.83.014703}. 
In this case, an antiferromagnetic order can be regarded as an odd-parity multipole order, accompanying both magnetic toroidal and quadrupole components~\cite{Spaldin:0953-8984-20-43-434203,kopaev2009toroidal}.
Accordingly, the antiferromagnetic order acquires an unusual magnetoelectric coupling to an electric current~\cite{Yanase:JPSJ.83.014703,Hayami_PhysRevB.90.024432}.
Therefore, such odd-parity multipoles are intriguing for exploring new spin-orbital coupled phenomena. 
However, the microscopic theory is not fully investigated. 
In particular, it is still unclear when and how such multipole orders are stabilized. 
It is highly desired to develop a general framework of the microscopic theory for the systems with local parity mixing. 

Experimentally, the local inversion symmetry breaking is widely found in multi-orbital systems on particular lattice structures. 
For instance, there are several $f$-electron materials possessing the lattice structures without local inversion symmetry, such as 
ferromagnetic superconductors UGe$_2$~\cite{oikawa1996crystal,saxena2000superconductivity,Huxley_PhysRevB.63.144519}, URhGe~\cite{aoki2001coexistence,levy2005magnetic,Hardy_PhysRevLett.94.247006}, and UCoGe~\cite{huy2007superconductivity,Huy_PhysRevLett.100.077002,aoki2009extremely}, the zig-zag chain compounds $LnM$$_2$Al$_{10}$ ($Ln$$=$Ce, Nd, Gd, Dy, Ho, and Er; $M$$=$Fe, Ru, and Os)~\cite{thiede1998ternary,reehuis2003magnetic,KhalyavinPhysRevB.82.100405,TanidaPhysRevB.84.115128,tanida2010possible,kondo2010magnetization,muro2011magnetic,mignot2011neutron}, the distorted honeycomb compounds $\alpha$- and $\beta$-YbAlB$_4$~\cite{macaluso2007crystal,nakatsuji2008superconductivity,matsumoto2011quantum}, and the diamond-structure compounds $R$$T$$_2$$X$$_{20}$ ($R$$=$Pr, La, Yb, and U, $T$$=$Fe, Co, Ti, V, Nb, Ru, Rh, and Ir, and $X=$Al and Zn)~\cite{moze1998crystal,Bauer_PhysRevB.74.155118,torikachvili2007six,onimaru2010superconductivity,sakai2011kondo,higashinaka2011single,onimaru2012nonmagnetic,onimaru2012simultaneous,matsubayashi2012pressure,sakai2012superconductivity,Tsujimoto_PhysRevLett.113.267001,ikeura2014anomalous}.
Although these materials can be candidates for unusual odd-parity multipole ordering through the local parity mixing, their properties have not been studied from such a viewpoint. 
In order to stimulate experiments and theories, it is important to clarify the microscopic mechanism for odd-parity multipole ordering. 

In this paper, we investigate a microscopic model by taking into account the 
hybridization between conduction and localized orbitals with different parity; e.g., between a $d$ component in conduction electrons and localized $f$ orbitals.
Starting from the periodic Anderson model with the antisymmetric hybridization defined on a quasi-one-dimensional lattice structure composed of zig-zag chains, we derive an effective low-energy model with antisymmetric exchange couplings between conduction electrons and localized spins. 
This is an extension of the Kondo lattice model to the case with local parity mixing.  
We show that the effective antisymmetric exchange couplings stabilize a 
N{\'e}el-type antiferromagnetic order with the spin polarization perpendicular to the zig-zag plane.
This is an odd-parity multipole order, which gives rise to a peculiar band deformation and magnetoelectric effects~\cite{Yanase:JPSJ.83.014703,Hayami_PhysRevB.90.024432}. 
Using several numerical calculations, we clarify that the multipole order is widely stabilized at and near half filling in the extended Kondo lattice model.

The organization of this paper is as follows. 
In Sect.~\ref{sec:Model}, we describe the derivation of the Kondo lattice model including antisymmetric exchange couplings from an extended periodic Anderson model. 
In Sect.~\ref{sec:results}, we clarify how the 
antisymmetric exchange couplings favor a multipole order associated with an antiferromagnetic order.
We also compute the electronic band structure and magnetoelectric effect under this multipole ordering. 
Moreover, we examine the stability of the multipole order systematically by the variational calculation for the ground state, simulated annealing, and Monte Carlo simulation at finite temperatures. 
The last section is devoted to a summary of this paper. 
The canonical transformation leading to the extended Kondo model is given in Appendix. 

\section{Model}
\label{sec:Model}

\subsection{Periodic Anderson model with antisymmetric hybridization}
\label{subsec:PAM-ASOC}

In this paper, we consider the effect of the antisymmetric spin-orbit coupling in the presence of local parity mixing. 
For this purpose, we introduce an extended periodic Anderson model with the antisymmetric hybridization between different parity orbitals~\cite{Yanase_doi:10.1143/JPSJ.77.124711}, whose Hamiltonian is given by 
\begin{align}
\label{eq:Ham_PAMSO}
\mathcal{H} &= \mathcal{H}_0 +\mathcal{H}_1, 
\end{align}
where 
\begin{align}
\label{eq:Ham_PAMSO1}
\mathcal{H}_0 
&= -\sum_{i,j,\sigma} 
 (t_{ij} c^{\dagger}_{i \sigma} c^{}_{j \sigma} + {\rm H.c.}) 
+ U \sum_i n_{i \uparrow}^f n_{i \downarrow}^f + E_0 \sum_{i, \sigma} n_{i \sigma}^f, \\
\label{eq:Ham_PAMSO2}
\mathcal{H}_1 
&= \sum_{l,\bm{k} ,\sigma} V_l({\bm{k}})
( c^{\dagger}_{l\bm{k} \sigma}f^{}_{l\bm{k} \sigma}+{\mathrm{H.c.}} ) 
+ \sum_{l,\bm{k},\sigma} 
\bm{g}^{cf}_l(\bm{k}) \cdot \bm{s}^{cf}_l (\bm{k}). 
\end{align}
Here, $c^{\dagger}_{i \sigma}$($c^{}_{i \sigma}$) and $f^{\dagger}_{i \sigma}$($f^{}_{i \sigma}$) 
are the creation (annihilation) operators of conduction and localized electrons with spin $\sigma$ at site $i=(p,l )$ ($p$ and $l$ denote the indices for the unit cell and sublattice, respectively); 
$c^{\dagger}_{l\bm{k} \sigma}$($c^{}_{l\bm{k} \sigma}$) and $f^{\dagger}_{l\bm{k} \sigma} (f^{}_{l\bm{k} \sigma})$ 
are their Fourier transformations, respectively. 
The first term in Eq.~(\ref{eq:Ham_PAMSO1}) represents the kinetic energy of conduction electrons; $t_{ij}$ is the hopping matrix element from site $j$ to $i$. 
The second and third terms are the on-site Coulomb interaction for localized electrons and the atomic energy of localized electrons, respectively ($n^f_{i\sigma} = f^{\dagger}_{i \sigma} f^{}_{i \sigma}$). 
Meanwhile, the first term in Eq.~(\ref{eq:Ham_PAMSO2}) represents the (symmetric) hybridization between conduction and localized electrons. 
We note that $\mathcal{H}_0$ and the first term in $\mathcal{H}_1$ comprise the conventional periodic Anderson model~\cite{Anderson_PhysRev.124.41}. 

On the other hand, the second term in Eq.~(\ref{eq:Ham_PAMSO2}) is introduced to describe an antisymmetric hybridization between $c$ and $f$ electrons; 
$\bm{g}_l^{cf} (\bm{k})$ is the polar vector describing the antisymmetric part, and 
\begin{align}
\bm{s}_l^{cf}(\bm{k}) 
= \sum_{\sigma, \sigma'} (c^{\dagger}_{l\bm{k} \sigma} \bm{\sigma}_{\sigma \sigma'} f^{}_{l\bm{k}\sigma'} 
+ {\rm H.c.}),  
\end{align} 
where $\bm{\sigma}=(\sigma^x, \sigma^y, \sigma^z)$ is the vector of Pauli matrices, as in Eq.~(\ref{eq:Ham_ASOC_def}). 
This term originates from the cooperation between the odd-parity crystalline electric field, atomic spin-orbit coupling, and off-site hybridization between orbitals with different parity~\cite{Yanase_kotai,Hayami_PhysRevB.90.024432}. 
Although similar hybridizations may appear between the same $c$-$c$ or $f$-$f$ orbitals, we omit them and consider only the hybridization between different orbitals $c$ and $f$ in the present model. 
This is justified in some realistic situations, e.g., when the atomic spin-orbit coupling for conduction electrons is weak and the intersite overlap integrals between localized electrons are small. 

The model in Eq.~(\ref{eq:Ham_PAMSO}) is applicable to the generic systems without spatial inversion symmetry.
In the present study, we focus on a specific situation where the antisymmetric hybridization originates from local parity mixing. 
As mentioned in the introduction, the local parity mixing exists in the lattice structures in which the inversion symmetry is broken at each site in a sublattice-dependent form; 
typical examples are a zig-zag chain, honeycomb lattice, and diamond lattice. 
On these lattices, the sublattice-dependent antisymmetric hybridization is present because each site is affected by the sublattice-dependent odd-parity crystalline electric field~\cite{Hayami_PhysRevB.90.024432}. 

\begin{figure}[htb!]
\begin{center}
\includegraphics[width=1.0 \hsize]{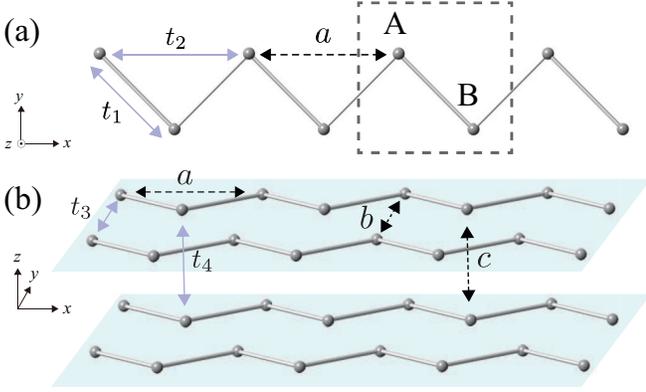} 
\caption{
\label{Fig:zigzag_chain}
Schematic pictures of (a) a single zig-zag chain and (b) a three-dimensional lattice composed of the zig-zag chains. 
The zig-zag chain is in the $xy$ plane: 
the $x$ and $y$ axes are taken in the parallel and perpendicular directions to the chain, respectively. 
The dashed rectangle represents the unit cell including two sublattice sites, A and B. 
$a$, $b$, and $c$ are the lattice constants; 
$t_1$, $t_2$, $t_3$, and $t_4$ represent the hopping integrals. 
}
\end{center}
\end{figure}

In the following sections, we consider one of the simplest realizations of the local parity mixing, a three-dimensional system composed of weakly-coupled one-dimensional zig-zag chains, as shown in Fig.~\ref{Fig:zigzag_chain}. 
We set the $x$ and $y$ axes in the plane on which the chains lie, with taking the $x$ axis in the chain direction;  
we set the lattice constants $a=b=c=1$. 
For this setting, the inversion symmetry is broken at each lattice site, and the odd-parity crystalline electric field is present along the $y$ direction with an alternating sign for two sublattices. 
Then, the local parity mixing results in the sublattice-dependent antisymmetric vector $\bm{g}^{cf}_{l}(\bm{k})$. 
For the current system with the quasi-one-dimensional structure, $\bm{g}_l^{cf}(\bm{k})$ is approximately given by
\begin{align}
\label{eq:gsimple}
\bm{g}_l^{cf}(\bm{k}) = 
(-1)^l g \hat{\bm{z}} \sin k_x, 
\end{align}
where $l=0(1)$ for the A(B) sublattice, $g$ is the coupling constant, and $\hat{\bm{z}}$ is the unit vector 
in the $z$ direction. 
The form of $\bm{g}_l^{cf}(\bm{k}) \parallel \hat{z}$ is understood from Eq. (1) by considering the quasi one-dimensional kinetic motion $\bm{k} \parallel \hat{x}$ and $\bm{\nabla}V_{\rm pot} \parallel \hat{y}$.
Note that $\bm{g}_l^{cf}(\bm{k})$ acts only between $c$ and $f$ electrons on the same sublattice.

\subsection{
Extended Kondo lattice model}
\label{subsec:generalized KLM}

With these preliminaries in the previous subsection, we derive an effective low-energy model by treating ${\mathcal H}_1$ as the perturbation to ${\mathcal H}_0$.
This is done by a standard procedure on the basis of the canonical transformation, that is the Schrieffer-Wolff transformation~\cite{Schrieffer_PhysRev.149.491}. 
Note that the procedure for the model without the antisymmetric hybridization term (i.e., $g=0$) leads to the standard Kondo lattice model~\cite{Schrieffer_PhysRev.149.491,Lacroix_PhysRevB.20.1969,Hewson199704}. 
In other words, we here extend the Kondo lattice model by including the effect of the antisymmetric hybridization. 
In the derivation, we assume that the number of $f$ electrons is one at each site, as in the standard procedure. 
We also assume
that the hybridization $V_l(\bm{k})$ has only the on-site component and sublattice independent: we drop the $\bm{k}$ and $l$ dependences.
By the straightforward calculations given in Appendix in detail, we obtain the effective Hamiltonian as 
\begin{align}
\label{eq:Ham_zigzag}
&\mathcal{H}_{{\rm ex-KLM}} = 
\sum_{i,j,\sigma} (t_{ij} c_{i\sigma}^{\dagger}c^{}_{j \sigma} + {\rm H.c.})
+ \frac{J}{2} \sum_{i,\sigma,\sigma'} c_{i \sigma}^{\dagger} \bm{\sigma}_{\sigma \sigma'} c^{}_{i \sigma'} \cdot \bm{S}_i\nonumber \\
&\qquad+
 \sum_{l,p, \bm{k}, \bm{k}'}
 \left\{ \frac{D_l}{2} 
 {\rm s}_{k_x} 
(S_{pl}^{+} s^-_{l\bm{k}'\bm{k}}
-S_{pl}^{-} s^+_{l\bm{k}'\bm{k}}
+S_{pl}^{z} n_{l\bm{k}'\bm{k}}) \delta_{\bm{R}_p} + {\rm H.c} \right\}
\nonumber \\ 
&\qquad+ \frac{G}{2}\sum_{l,p, \bm{k}, \bm{k}'} 
{\rm s}_{k_x} {\rm s}_{k'_x} (2S_{pl}^{z} s^z_{l\bm{k}'\bm{k}}-S_{pl}^{+} s^-_{l\bm{k}'\bm{k}}
-S_{pl}^{-} s^+_{l\bm{k}'\bm{k}}) \delta_{\bm{R}_p}, 
\end{align}
where 
${\rm s}_k = \sin k$, 
$\delta_{\bm{R}_p} = e^{-{\rm i}(\bm{k}'-\bm{k})\cdot \bm{R}_{p}}$, 
$s^+_{l\bm{k}'\bm{k}} = c_{l\bm{k}'\uparrow}^{\dagger} c^{}_{l\bm{k} \downarrow}$, 
$s^-_{l\bm{k}'\bm{k}} = c_{l\bm{k}'\downarrow}^{\dagger} c^{}_{l\bm{k} \uparrow}$, 
$s^z_{l\bm{k}'\bm{k}} = (c_{l\bm{k}'\uparrow}^{\dagger} c^{}_{l\bm{k} \uparrow}-c_{l\bm{k}'\downarrow}^{\dagger} c^{}_{l\bm{k} \downarrow})/2$, and 
$n_{l\bm{k}'\bm{k}} = c_{l\bm{k}'\uparrow}^{\dagger} c^{}_{l\bm{k} \uparrow}+c_{l\bm{k}'\downarrow}^{\dagger} c^{}_{l\bm{k} \downarrow}$; 
$\bm{S}_{pl}=(S^x_{pl},S^y_{pl},S^z_{pl})$ represents the localized spins at the unit cell $p$ and sublattice $l$ and $S^{\pm}_{pl}=S^x_{pl} \pm {\rm i} S^y_{pl}$. 

In Eq.~(\ref{eq:Ham_zigzag}), the first and second terms are the kinetic energy of conduction electrons and the on-site exchange coupling between localized spins and conduction electrons, respectively. 
These two terms represent the standard Kondo lattice model. 
Here, $J$ is in the second order of $V$, as shown in Eqs.~(\ref{eq:J_expression}) and (\ref{eq:J_simple}). 
In the first term, we take into account four different hopping matrix elements: intrachain hoppings between nearest and next-nearest neighbor sites, $t_1$ and 
$t_2$, respectively [see Fig.~\ref{Fig:zigzag_chain}(a)], and interchain hoppings between the same sublattices for the neighboring sites in the $y$ and $z$ directions, $t_3$ and $t_4$, respectively 
[see Fig.~\ref{Fig:zigzag_chain}(b)]. 

The third and fourth terms in Eq.~(\ref{eq:Ham_zigzag}) are the antisymmetric exchange couplings, derived from the antisymmetric hybridization in Eq.~(\ref{eq:Ham_PAMSO2}). 
The third term results from the combination of $V$ and $\bm{g}_l^{cf}(\bm{k})$. 
The coefficient $D_l$ is given by 
\begin{align}
D_l= 
(-1)^l \sqrt{JG}, 
\end{align}
where $G$ is proportional to $g^2$ [see Eqs.~(\ref{eq:G_expression}) and (\ref{eq:G_simple})]: 
$D_l$ is proportional to $gV$. 
The $s_{k_{x}}=\sin k_x$ dependence in Eq.~(\ref{eq:Ham_zigzag}) comes from Eq.~(\ref{eq:gsimple}), which is the source of the band deformation and magnetoelectric effect, similar to the case of toroidal ordering~\cite{Yanase:JPSJ.83.014703,Hayami_PhysRevB.90.024432}; 
this is demonstrated in Sect.~\ref{sec:Electronic Structure}. 
The fourth term in Eq.~(\ref{eq:Ham_zigzag}) comes from the second order of $\bm{g}_l^{cf}(\bm{k})$.  
It plays no essential role in the peculiar electronic and transport properties, since it is symmetric with respect to $(\bm{k},\bm{k})\to (-\bm{k},-\bm{k}')$.
Hereafter, assuming the situation $V>g$, we mainly consider the case with $J>|D_l|>G$; we take $J$ and $G$ to be positive. 
For simplicity, we treat $\bm{S}_i$ as a classical spin with $|\bm{S}_i|=1/2$ in the following analysis.

\section{Results}
\label{sec:results}

\subsection{Two-site problem}
\label{subsec:2site}

\begin{figure}[htb!]
\begin{center}
\includegraphics[width=1.0 \hsize]{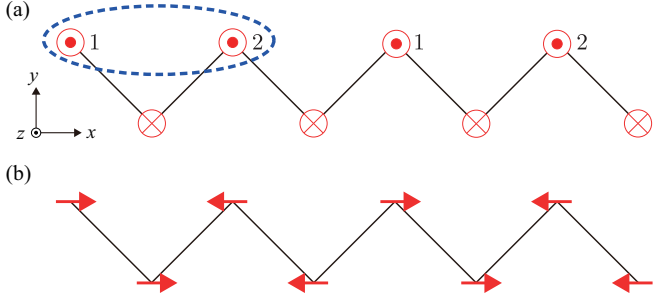} 
\caption{
\label{Fig:zigzag_spin_ponti}
Schematic pictures of magnetic orders on the zig-zag chain: (a) the antiferromagnetic order with up-down spin moments along the $z$ direction ($z$-UD) and (b) the up-up-down-down antiferromagnetic order with the moments along the $x$ direction ($x$-UUDD). 
In (a), the dashed oval shows the two neighboring sites in the same sublattice, $1$ and $2$. 
}
\end{center}
\end{figure}

Let us first examine the effect of the third and fourth terms in Eq.~(\ref{eq:Ham_zigzag}) by considering a simple two-site problem.
We discuss what type of spin configuration is stabilized by these antisymmetric exchange couplings arising from the antisymmetric hybridization. 
We consider two sites connected by $t_2$ in a single chain, $1$ and $2$ (belonging to the same sublattice, say A-sublattice), as shown in Fig.~\ref{Fig:zigzag_spin_ponti}(a).
For the two sites, the exchange couplings in Eq.~(\ref{eq:Ham_zigzag}) are explicitly written in the matrix form as
\begin{align}
\label{eq:2site_mat}
\tilde{\mathcal{H}}_{
2 \, {\rm site}} = 
\left(
\begin{array}{cc}
\tilde{\mathcal{H}}^{11}_{2 \, {\rm site}}
&
\tilde{\mathcal{H}}^{12}_{2 \, {\rm site}} \\
\tilde{\mathcal{H}}^{21}_{2 \, {\rm site}}
&
\tilde{\mathcal{H}}^{22}_{2 \, {\rm site}}
\end{array}
\right), 
\end{align}
where  
\begin{align}
\tilde{\mathcal{H}}^{11}_{2 \, {\rm site}} &= \left( \frac{J}{2} S^{z}_{1}+\frac{G}{2} {\rm s}_{k_x}^2 S^z_{2} \right) \sigma^z 
+ \left( \frac{J}{2}S_{1}^x -\frac{G}{2} {\rm s}_{k_x}^2 S^x_{2}  \right)  \sigma^x, \\
\tilde{\mathcal{H}}^{12}_{2 \, {\rm site}} &  =\frac{\sqrt{JG}}{2} {\rm s}_{k_x} (S^z_{1}+S^z_{2}) \sigma^0 
 + {\rm i} \frac{\sqrt{JG}}{2} {\rm s}_{k_x} (S_{1}^x-S_{2}^x) \sigma^y. 
 \label{eq:2site_mat_offdiag}
\end{align}
$\tilde{\mathcal{H}}^{21}_{2 \, {\rm site}}=\tilde{\mathcal{H}}^{12\,\dagger}_{2 \, {\rm site}}$ and $\tilde{\mathcal{H}}^{22}_{2 \, {\rm site}}$ is given by $\tilde{\mathcal{H}}^{11}_{2 \, {\rm site}}$ with $1 \leftrightarrow 2$.  
We take the basis $(c_{1 \uparrow}, c_{1 \downarrow}, c_{2 \uparrow}, c_{2 \downarrow})$ in Eq.~(\ref{eq:2site_mat}), and $\sigma^0$ is the $2 \times 2$ unit matrix.  
Here, we omit the $y$ component of localized spins, $S^y$, without loss of generality, as it gives an equivalent contribution to that from $S^x$ due to the spin rotational symmetry in the $xy$ plane.
Equation~(\ref{eq:2site_mat}) indicates that 
there is an effective hopping of conduction electrons, induced by the antisymmetric exchange coupling proportional to $\sqrt{JG}$. 
The effective hopping depends on the directions of localized spins, as shown in Eq.~(\ref{eq:2site_mat_offdiag}).
This differentiates the energies for different spin configurations. 

We first assume two different types of ferromagnetic configurations for the two A-sublattice sites: one is the state with spin polarization in the $x$ direction, and the other in the $z$ direction. 
For these two cases, the matrix elements of the Hamiltonian in Eq.~(\ref{eq:2site_mat}) are given by
\begin{align}
\tilde{\mathcal{H}}^{x{\rm -F}}_{
2 \, {\rm site}} &=\left( 
\begin{array}{cc}\displaystyle
\left(\frac{J}{4}-\frac{G}{4} 
{\rm s}_{k_x}^2 \right) \sigma^x
& 0  \\ 
0
&\displaystyle \left( \frac{J}{4}-\frac{G}{4} 
 {\rm s}_{k_x}^2 \right) \sigma^x  \\
\end{array}
\right), 
\\
\tilde{\mathcal{H}}^{z{\rm -F}}_{2 \, {\rm site}} &=\left( 
\begin{array}{cc}\displaystyle
\left( \frac{J}{4} + \frac{G}{4} {\rm s}_{k_x}^2 \right) \sigma^z
 &\displaystyle  -\frac{\sqrt{JG}}{2} 
 {\rm s}_{k_x}  \sigma^0  \\\displaystyle
  -\frac{\sqrt{JG}}{2} {\rm s}_{k_x}  \sigma^0
 &\displaystyle \left( \frac{J}{4}+\frac{G}{4} 
 {\rm s}_{k_x}^2 \right) \sigma^z \\
\end{array}
\right). 
\end{align}
The eigenvalues are 
obtained as 
\begin{align}
\label{eq:energy-Fx-A-sublattice}
\varepsilon^{x{\rm -F}}_{
2 \, {\rm site}} (k_x) &= 
\pm \left( \frac{J}{4}- \frac{G}{4}{\rm s}_{k_x}^2 \right), \\
\label{eq:energy-Fz-A-sublattice}
\varepsilon^{z{\rm -F}}_{
2 \, {\rm site}} (k_x) &=\pm \left( \frac{J}{4} + \frac{G}{4} {\rm s}_{k_x}^2\right) \pm \frac{\sqrt{JG}}{2} {\rm s}_{k_x}.
\end{align}
The results in Eqs.~(\ref{eq:energy-Fx-A-sublattice}) and~(\ref{eq:energy-Fz-A-sublattice}) clearly show that the antisymmetric exchange couplings prefer to the ferromagnetic configuration with magnetic moments in the $z$ direction for the two A-sublattice sites. 

Next, let us consider a N\'eel-type antiferromagnetic spin configuration for the A-sublattice sites. 
In this case, the Hamiltonian is given by 
\begin{align}
\tilde{\mathcal{H}}^{x{\rm -AF}}_{
2 \, {\rm site}} &=\left(
\begin{array}{cc}\displaystyle
\left(  \frac{J}{4}+\frac{G}{4} {\rm s}^2_{k_x} \right) \sigma^x
 &\displaystyle  {\rm i} \frac{\sqrt{JG}}{2} {\rm s}_{k_x} \sigma^y \\\displaystyle
- {\rm i} \frac{\sqrt{JG}}{2} {\rm s}_{k_x} \sigma^y
 &\displaystyle -\left(  \frac{J}{4}+\frac{G}{4} {\rm s}^2_{k_x} \right) \sigma^x
\end{array}
\right),  \\
\tilde{\mathcal{H}}^{z{\rm -AF}}_{
2 \, {\rm site}} &=\left(
\begin{array}{cc}\displaystyle
\left( \frac{J}{4}-\frac{G}{4} {\rm s}^2_{k_x} \right) \sigma^z
 & 0  \\
 0
 &\displaystyle  -\left( \frac{J}{4}-\frac{G}{4} {\rm s}^2_{k_x} \right) \sigma^z
 \end{array}
\right),  
\end{align}
for the cases with the antiferromagnetic moments along the $x$ and $z$ directions, respectively. 
The eigenvalues are obtained as 
\begin{align}
\label{eq:energy-AFx-A-sublattice}
\varepsilon^{x{\rm -AF}}_{
2 \, {\rm site}} (k_x) &=  
\pm \left( \frac{J}{4} +\frac{G}{4} {\rm s}^2_{k_x} \right) \pm \frac{\sqrt{JG}}{2} {\rm s}_{k_x}, \\
\label{eq:energy-AFz-A-sublattice}
\varepsilon^{z{\rm -AF}}_{
2 \, {\rm site}} (k_x) &= 
\pm \left( \frac{J}{4}-\frac{G}{4} {\rm s}^2_{k_x} \right).
\end{align}
Thus, the stability of the $x$ and $z$ orders is opposite to the ferromagnetic case: 
the antisymmetric exchange couplings favor the antiferromagnetic configuration with magnetic moments in the $x$ direction. 

On the other hand, the spin configuration between the A and B sublattices will depend on an effective exchange coupling induced by the intersublattice hopping $t_1$. 
For instance, when $t_1$ is much larger than $t_2$, the antiferromagnetic configuration between two sublattices is induced at and near half filling in the strongly-correlated region. 
This is partly understood by considering the second-order perturbation with respect to $t_1$ at half filling, which leads to an effective antiferromagnetic interaction between localized spins.  
Meanwhile, the ferromagnetic configuration is favored at low and high fillings due to the double-exchange mechanism, which is an effective ferromagnetic interaction between localized spins induced by the kinetic motion of itinerant electrons~\cite{Zener_PhysRev.82.403,anderson1955considerations}. 

The above considerations suggest that the antisymmetric exchange couplings in Eq.~(\ref{eq:Ham_zigzag}) stabilize a specific spin configuration with a preference on the direction of magnetic moments. 
For the ferromagnetic configuration in the same sublattice, the $z$-component order is preferred, which results in the up-down type antiferromagnetic state ($z$-UD) at and near half filling [see Fig.~\ref{Fig:zigzag_spin_ponti}(a)], and the ferromagnetic state ($z$-F) at low and high fillings. 
On the other hand, for the antiferromagnetic configuration for the same sublattice, the $x$-component order is favored, leading to the up-up-down-down type antiferromagnetic state ($x$-UUDD) irrespective of the sign in the effective exchange coupling between different sublattices[see Fig.~\ref{Fig:zigzag_spin_ponti}(b)]. 

In the following sections, we focus on the $z$-UD order, since it accompanies an odd-parity multipole order composed of magnetic toroidal and quadrupole components~\cite{Yanase:JPSJ.83.014703}.
In this case, the ferroic toroidal order $\bm{T}$ is induced in the $x$ direction: 
$\bm{T}\propto (\bm{\nabla}_{l}V_{{\rm pot}})\times\langle\bm{S}_{l}\rangle$.
In the following calculations, for simplicity, we restrict ourselves to the ordered states with the ordering vector, $\bm{q}=(q_{x},0,0)$.

\subsection{Band deformation and magnetoelectric effect}
\label{sec:Electronic Structure}

In this subsection, we examine the nature of the $z$-UD ordered state in Fig.~\ref{Fig:zigzag_spin_ponti}(a). 
As shown in the previous study~\cite{Yanase:JPSJ.83.014703}, the $z$-UD order is expected to cause a band deformation with a shift of the band bottom. 
Also, the system may exhibit the magnetoelectric effect in which a staggered magnetic moment is induced by an electric current.  
We note that the magnetoelectric effect takes place even in the paramagnetic state, due to the presence of the antisymmetric exchange couplings.
We here confirm that these effects occur also in our extended Kondo lattice model in Eq.~(\ref{eq:Ham_zigzag}). 

\begin{figure}[htb!]
\begin{center}
\includegraphics[width=0.8 \hsize]{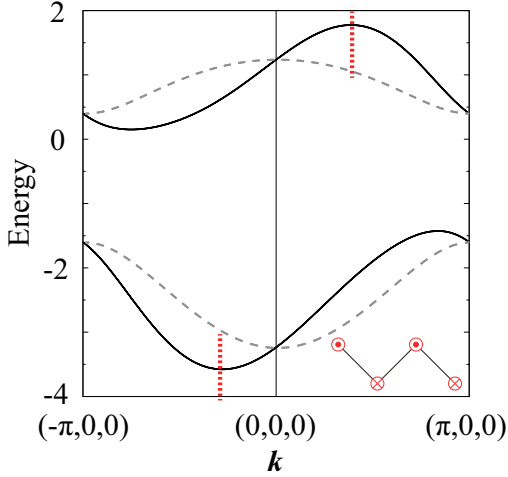} 
\caption{
\label{Fig:band}
Energy dispersion along the $k_x$ direction of the Hamiltonian in Eq.~(\ref{eq:Ham_zigzag}) under the $z$-UD order, as shown in the inset. 
Each band is doubly degenerate. 
The result is calculated at $t_1=1$, $t_2=0.1$, $t_3=0.2$, $t_4=0.2$, $J=4$, and $G=0.5$. 
The dotted red lines indicate the band bottoms. 
The dashed black curves represent the energy dispersion at $G=0$.
}
\end{center}
\end{figure}

Figure~\ref{Fig:band} shows the band structure in our model in Eq.~(\ref{eq:Ham_zigzag}) 
along the $k_x$ direction from $(-\pi,0,0)$ to $(\pi,0,0)$. 
The result is calculated by assuming the presence of the $z$-UD type antiferromagnetic order with full polarization in localized electrons: $\bm{S}_{pl} = (-1)^l(\hat{\bm{z}}/2)$. 
We take $t_1=1$, $t_2=0.1$, $t_3=0.2$, $t_4=0.2$, $J=4$, and $G=0.5$. 
We also show the result at $G=0$ for comparison.
As shown in Fig.~\ref{Fig:band}, 
the $z$-UD order in the presence of the antisymmetric exchange couplings modifies the band structure in a peculiar way, with shifting the band bottom in the $k_x$ direction. 
This is similar to the previous results~\cite{Yanase:JPSJ.83.014703,Hayami_PhysRevB.90.024432}: 
although both spatial-inversion $\mathcal{P}$ and time-reversal $\mathcal{T}$ symmetries are broken in this ordered state, 
the combined $\mathcal{PT}$ symmetry is retained, which ensures $E_\sigma(\bm{k}) = E_{-\sigma}(\bm{k})$ (two-fold degeneracy in each band), while there is no guarantee to satisfy $E_\sigma(\bm{k}) = E_{\sigma}(-\bm{k})$.  

In our extended Kondo lattice model in Eq.~(\ref{eq:Ham_zigzag}), 
this band deformation is understood as follows. 
The Hamiltonian for the $z$-UD ordered state is represented by 
\begin{align}
\label{eq:Ham_chain}
\tilde{\mathcal{H}}_{{\rm chain}} =\left(
\begin{array}{cc}
u+v
 &  \varepsilon_{{\rm 1}} \\
\varepsilon_{{\rm 1}}
 &  u-v
 \\
\end{array}
\right),  
\end{align}
where 
$u = \varepsilon_{{\rm 2}} + G\sigma {\rm s}^2_{k_x}/4$ and 
$v = (J\sigma/2 + \sqrt{JG} {\rm s}_{k_x} )/2$ ($\sigma = \pm 1$ for up and down spins); 
$\varepsilon_1= - 2 t_1 \cos (k_x/2)$ and $\varepsilon_2 = -2t_2 \cos k_x -2 t_3 \cos k_y -2 t_4 \cos k_z$ represent the energy dispersions for the different and same sublattices, respectively.
We take the basis as $(c_{{\rm A}\bm{k}\sigma}, c_{{\rm B}\bm{k}\sigma})$, where $c_{{\rm A}\bm{k} \sigma}$($c_{{\rm B}\bm{k} \sigma}$) is the annihilation operator at A(B) sublattice with wave vector $\bm{k}$ and spin $\sigma$. 
By diagonalizing the Hamiltonian in Eq.~(\ref{eq:Ham_chain}), the eigenvalues are obtained as 
\begin{align}
\label{eq:energy_updown}
E (\bm{k}) &= \varepsilon_2 + \frac{G}{4} {\rm s}^{2}_{k_x} 
\pm \sqrt{\varepsilon_1^2 + \frac14 \left(\frac{J}{2} + \sqrt{JG} {\rm s}_{k_x}\right)^2}. 
\end{align}
This indicates that the band deformation with the band bottom shift is due to $\sqrt{JG} {\rm s}_{k_x}$, which comes from the $ D_{l}{\rm s}_{k_x}$ term in Eq.~(\ref{eq:Ham_zigzag}).   
Equation~(\ref{eq:energy_updown}) also indicates that the peculiar band deformation does not occur for $G=0$ (the standard Kondo lattice model), as shown in Fig.~\ref{Fig:band}.

\begin{figure}[htb!]
\begin{center}
\includegraphics[width=0.85 \hsize]{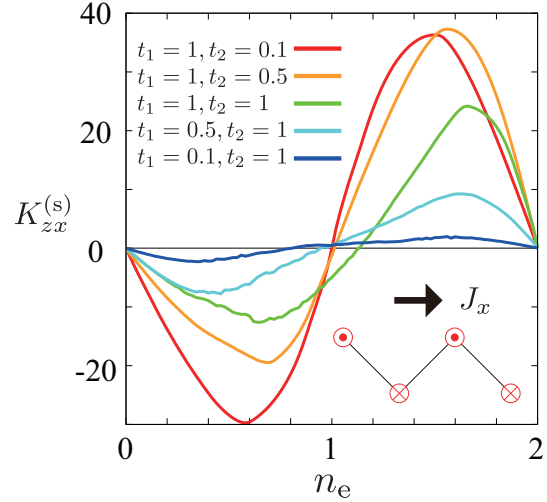} 
\caption{
\label{Fig:ME}
Correlation between the $z$-UD moment and electric current in the $x$ direction, $K_{zx}^{\rm (s)}$, in Eq.~(\ref{Eq:Magnetoelectric effect_sec3}). 
The data are obtained for several values of $t_1$ and $t_2$ indicated in the figure, at $t_3=0.2$, $t_4=0.2$, $J=4$, $G=0.5$, $T=0.01$, and $\eta=0.01$. 
The inset shows a schematic picture for the magnetoelectric response expected from the correlation $K_{zx}^{\rm (s)}$. 
}
\end{center}
\end{figure}

Next, let us discuss the magnetoelectric effect by computing the linear response of the staggered magnetization in the form of $z$-UD order by an electric current in the $x$ direction.
The response is calculated by the Kubo formula in the form: 
\begin{align}
K_{zx}^{\rm (s)} =\frac{g \mu_{{\rm B}}}{2}\frac{1}{{\rm i} V_{{\rm s}}} \sum_{m,n,\bm{k}} \frac{f(\varepsilon_{n \bm{k}})-f(\varepsilon_{m \bm{k}})}{\varepsilon_{n \bm{k}}-\varepsilon_{m \bm{k}}} 
\frac{
\sigma_{z,\bm{k}}^{nm} J_{x,\bm{k}}^{mn} }{
\varepsilon_{n \bm{k}}-\varepsilon_{m \bm{k}}+{\rm i} \eta}, 
\label{Eq:Magnetoelectric effect_sec3}
\end{align}
where $V_{{\rm s}}$ is the system volume, $f(\varepsilon)$ is the Fermi distribution function, $\sigma_{z,\bm{k}}^{nm}=\langle n \bm{k} | (-1)^l \sigma^{z} | m\bm{k} \rangle$, and $J_{x,\bm{k}}^{mn}=\langle m \bm{k} |J_{x} | n\bm{k} \rangle$ is the matrix element of the current operator
$J_x = (e/\hbar)\partial \mathcal{H}_{{\rm ex-KLM}}
/\partial k_x$; 
$\varepsilon_{n \bm{k}}$ and $| n \bm{k} \rangle$ are the $n$-th eigenvalue and eigenstate of $\mathcal{H}_{{\rm ex-KLM}}$ by assuming the $z$-UD order with full polarization in the localized spins. 
We set $g \mu_{{\rm B}} e/2h=1$. 
Figure~\ref{Fig:ME} shows the result of $K_{zx}^{{\rm (s)}}$ as a function of the electron density $n_{{\rm e}}=(1/N)\sum_{i \sigma} \langle c^{\dagger}_{i \sigma} c^{}_{i \sigma} \rangle$, where $N$ is the total number of sites. 
The results are calculated at $t_3=0.2$, $t_4=0.2$, $J=4$, $G=0.5$, while changing $t_1$ and $t_2$, which may correspond to a deformation of the zig-zag chain (see below).
We set $T=0.01$ by taking the damping factor $\eta=0.01$.
As shown in the inset of Fig.~\ref{Fig:ME}, the 
staggered moments in the $z$ direction are induced by the electric current in the $x$ direction, consistent with the previous study~\cite{Yanase:JPSJ.83.014703}. 
This is due to the presence of the toroidal component $T^{x}$ in the multipole order~\cite{Hayami_PhysRevB.90.024432}.

As shown in Fig.~\ref{Fig:ME}, the magnetoelectric response tends to be larger for larger $t_1/t_2$. 
This tendency is clearly seen in the low density region where the Fermi surface has a simple shape. 
This suggests that the shallower zig-zag structure, where 
the intersublattice hopping becomes dominant, gives rise to a larger magnetoelectric effect. 
It also implies that the magnetoelectric response can be controlled by a uniaxial pressure along or perpendicular to the zig-zag chain. 
The results are consistent with those in the single-band Hubbard model including the effect of the antisymmetric spin-orbit coupling~\cite{Yanase:JPSJ.83.014703}. 

Moreover, the response also depends on $J$ and $G$ in a peculiar manner (not shown here). 
The $G$ dependence is rather simple, at least, in the low density region. 
This is because the increase of $G$ results in the increment of the antisymmetric spin-orbit coupling as in the third term of Eq.~(\ref{eq:Ham_zigzag}), which is the origin of the staggered magnetoelectric response. 
Meanwhile, the $J$ dependence is complicated because the increase of $J$ enhances not only the antisymmetric spin-orbit coupling but also the exchange coupling as in the second term of Eq.~(\ref{eq:Ham_zigzag}).

\subsection{Phase diagram}
\label{subsec:phase diagram}

In the previous subsection, we have studied the electronic and transport properties in the assumed $z$-UD ordered state. 
Now, we examine when and how such an ordered state is realized in the model in Eq.~(\ref{eq:Ham_zigzag}). 
For that purpose, we perform three numerical calculations which are complementary to each other: variational calculation for the ground state in Sect.~\ref{sec:Variational Calculation}, simulated annealing in Sect.~\ref{sec:Simulated Annealing}, and Monte Carlo simulation at finite temperatures in Sect.~\ref{sec:Monte Carlo Simulation}.

\subsubsection{Variational calculation}
\label{sec:Variational Calculation}

\begin{figure}[htb!]
\begin{center}
\includegraphics[width=1.0 \hsize]{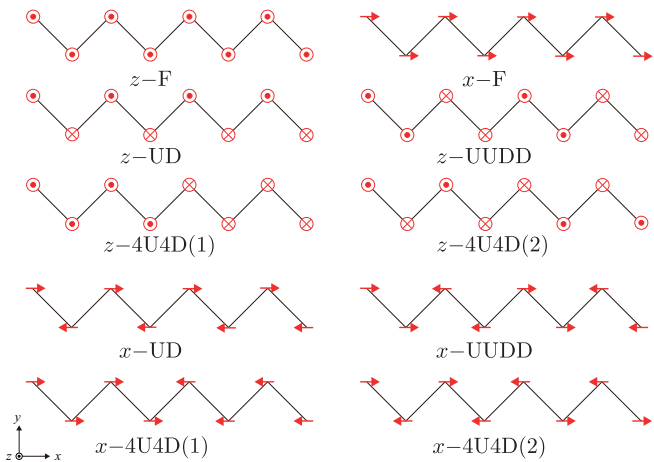} 
\caption{
\label{Fig:zigzag_variational}
Schematic pictures of ordering patterns of localized spins used in the variational calculations. 
U and D represent up- and down-spin polarization, respectively, and F represents the ferromagnetic state. 
The prefices $x$ and $z$ indicate the direction of the magnetic moments. 
}
\end{center}
\end{figure}

First, we examine the ground state of the model in Eq.~(\ref{eq:Ham_zigzag}) by 
a variational calculation. 
Namely, for each parameter set of the Hamiltonian, we compare the zero-temperature grand potential per site, $\Omega = E - \mu n_{{\rm e}}$ ($E = \langle \mathcal{H}_{{\rm ex-KLM}} \rangle /N$ is the internal energy per site and $\mu$ is the chemical potential) for different magnetically-ordered states and determine the most stable one which gives the lowest $\Omega$.
In the present calculations, we assume a collection of typical magnetic orders in the localized spins, up to the eight-site unit cell in the single chain as candidates for the ground state; 
the states considered in this study are shown in Fig.~\ref{Fig:zigzag_variational}. 
We consider only uniform $\bm{q}=\bm{0}$ orders for all these patterns; namely, we assume that each magnetically ordered pattern is composed of a uniform arrangement of the magnetic unit cell in all the directions. 
The data are computed by approximating the integral over the folded Brillouin zone using the sum over grid points of $64 \times 64 \times 64$. 
The phase separated regions are determined by the method in Ref.~\citen{Hayami_Conference}. 

\begin{figure}[htb!]
\begin{center}
\includegraphics[width=0.9 \hsize]{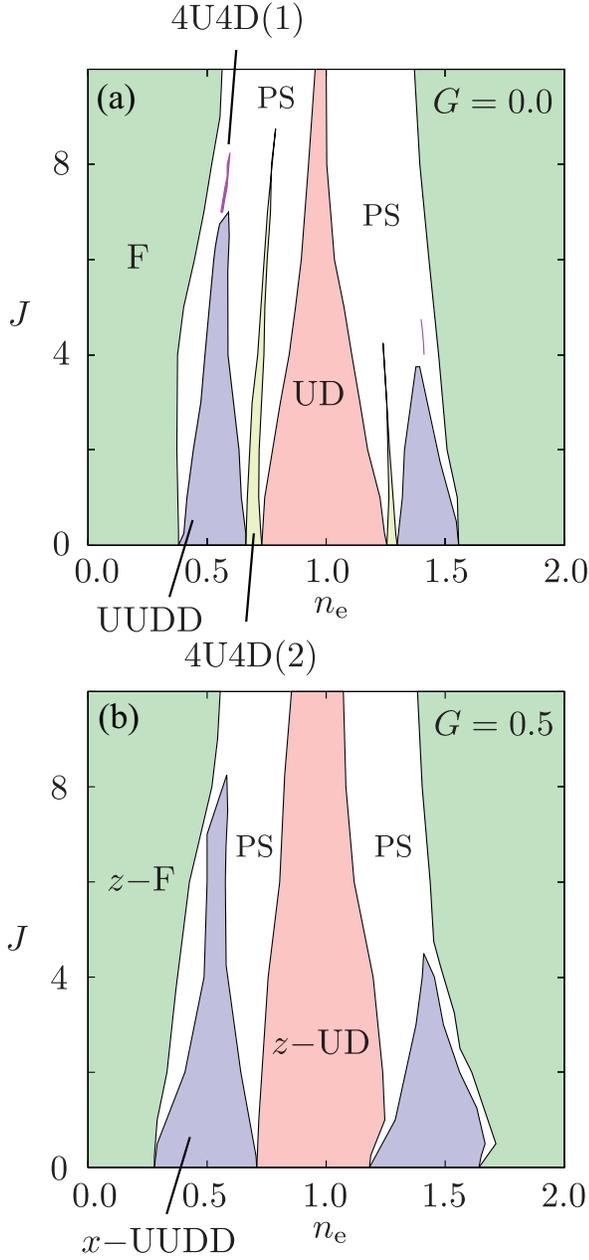} 
\caption{
\label{Fig:t1=1.0_t2=0.1_G=0_t3=0.2_souzu}
Ground-state phase diagram of the model in Eq.~(\ref{eq:Ham_zigzag}) obtained by the variational calculation at (a) $G=0$ and (b) $G=0.5$. 
Other parameters are taken to be $t_1=1$, $t_2=0.1$, $t_3=0.2$, and $t_4=0.2$. 
PS indicates a phase separated region. 
Ordering patterns are shown in Fig.~\ref{Fig:zigzag_variational}. 
}
\end{center}
\end{figure}

Figure~\ref{Fig:t1=1.0_t2=0.1_G=0_t3=0.2_souzu} shows the phase diagram obtained by the variational calculation, as a function of $n_{{\rm e}}$ and $J$. 
Figure~\ref{Fig:t1=1.0_t2=0.1_G=0_t3=0.2_souzu}(a) corresponds to the result for the standard Kondo lattice model 
without the antisymmetric exchange couplings, i.e., at $G=0$. 
In this case, there is no spin anisotropy because the model in Eq.~(\ref{eq:Ham_zigzag}) retains the spin rotational symmetry. 
In the low and high density regions, the ferromagnetic metallic phase appears and becomes wider as $J$ increases. 
The ferromagnetic phase is stabilized by the double-exchange mechanism~\cite{anderson1955considerations}.   
On the other hand, a staggered antiferromagnetic order along the chain is stabilized at and near half-filling ($n_{{\rm e}}=1$) due to the effective antiferromagnetic interaction mentioned in Sect.~\ref{subsec:2site}. 
Note that an incommensurate order might take over the antiferromagnetic state in the weak-coupling region, which cannot be described in the present variational scheme within the limited sizes of magnetic unit cells; we will reexamine this point in the following subsections. 
In the intermediate $n_{{\rm e}}$ region, there are several phases characterized by other ordering wave vectors: UUDD and 4U4D antiferromagnetic orders (see Fig.~\ref{Fig:zigzag_variational}).  
Note that, in all these phases, the common direction of the magnetic moments can be taken arbitrarily owing to spin rotational symmetry of the system. 

Next, we discuss the effect of the antisymmetric exchange couplings by turning on $G$. 
When $G\neq 0$, the antisymmetric exchange couplings introduce the spin anisotropy, and hence, the system has a preference in the direction of the magnetic moments in each phase. 
Figure~\ref{Fig:t1=1.0_t2=0.1_G=0_t3=0.2_souzu}(b) shows the result at $G=0.5$. 
At and near half filling, the quantized axis in the up-down antiferromagnetic phase prefers to be fixed in the $z$ direction. 
This is the $z$-UD phase with multipole ordering discussed in Sect.~\ref{sec:Electronic Structure}, which gives rise to a band deformation with a band bottom shift and exhibits the magnetoelectric effect. 
The result in Fig.~\ref{Fig:t1=1.0_t2=0.1_G=0_t3=0.2_souzu} indicates that this UD phase tends to be stabilized by $G$, as was discussed in Sect.~\ref{subsec:2site}. 

Meanwhile, the UUDD phases near quarter and three quarter fillings are also present at $G=0.5$, as shown in Fig.~\ref{Fig:t1=1.0_t2=0.1_G=0_t3=0.2_souzu}(b). 
In this case, however, the magnetic moments are aligned within the $xy$ plane by the spin anisotropy; 
we denote it by UUDD with the prefix $x$, while the moment can point to any direction within the $xy$ plane.
On the other hand, in the ferromagnetic phases in the low and high filling regions, the moments are polarized in the $z$ direction.

For all the phases we obtained, 
the direction of the magnetic moments as well as the stable parameter region is consistent with the arguments for the two-site problem discussed in Sect.~\ref{subsec:2site}. 
The tendency does not change while changing $t_1/t_2$ (not shown).

\subsubsection{Simulated annealing}
\label{sec:Simulated Annealing}

In order to confirm the stability of the $z$-UD state by a more unbiased method than the variational calculation in Sect.~\ref{sec:Variational Calculation}, we adopt simulated annealing. 
The simulated annealing is an optimization method for finding the global minimum of a function that possesses many local minima~\cite{kirkpatrick1983optimization}.  
By using the method, we obtain the accurate magnetic order in the ground state within the unit cell we set in the calculation; we do not need to assume a specific magnetic order, 
in contrast to the variational calculation in Sect.~\ref{sec:Variational Calculation}. 
We optimize the classical localized spins by the simulated annealing as follows: 
temperature $T$ is decreased gradually, and for each $T$, the spin configuration is updated by the Monte Carlo sampling with the single-spin flip algorithm.

In the present calculations, we performed the simulated annealing while decreasing $T$ in a geometrical way, $T_{n+1} = \alpha T_n$, where $T_n$ is the temperature in the $n$-th step. 
We started from the initial temperature $T_0=1.0$ by taking the coefficient of geometrical cooling $\alpha = 0.97$ and the total steps of cooling 270: the final temperature $T_{270}$ reaches down to $\sim 3.0\times 10^{-4}$. 
We considered the systems with $N=16 \times 1 \times 1$ and $24 \times 1 \times 1$ while introducing a supercell consisting of $N_k = 8^3$ copies of the $N$-site lattice to reduce the finite-size effect. 

\begin{figure}[htb!]
\begin{center}
\includegraphics[width=0.8 \hsize]{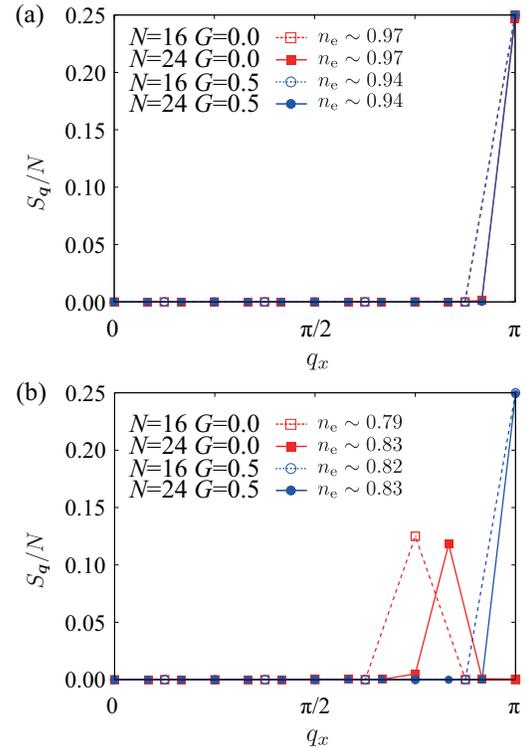} 
\caption{
\label{Fig:annealing_Kondo_zigzag}
$q_x$ dependence of the spin structure factor at $q_y=q_z=0$ divided by the system size $N$ obtained by the simulated annealing. 
See Ref.\citen{comment_wave_number}.
The data are calculated at $t_1=1$, $t_2=0.1$, $t_3=0.2$, $t_4=0.2$, and $J=4$. 
The results for $G=0$ and $G=0.5$ with the system size $N=16$ and 24 are shown in each figure: (a) $\mu=-0.4$ and (b) $\mu=-0.8$. 
Electron fillings for each $\mu$ are also shown in the figure.
}
\end{center}
\end{figure}

Figure~\ref{Fig:annealing_Kondo_zigzag} shows the spin structure factor obtained by the simulated annealing. 
The spin structure factor is defined by 
\begin{align}
\label{eq:Sq_sec3}
S_{\bm{q}} = \frac{1}{N} \sum_{i,j}  (\bm{S}_i \cdot \bm{S}_j)  e^{{\rm i} \bm{q}\cdot (\bm{r}_i - \bm{r}_j)}, 
\end{align}
where $\bm{q}$ is a wave vector and $\bm{r}_i$ is the position vector at site $i$. 
In Fig.~\ref{Fig:annealing_Kondo_zigzag}, $S_{\bm{q}}/N$ is plotted as a function of $q_x$ at $\bm{q} = (q_x,0,0)$. 
Note that, for the perfect UD order, $S_{\bm{Q}}/N = 0.25$ at $\bm{Q}=(\pi, 0, 0)$, and otherwise zero~\cite{comment_wave_number}. 
At $\mu=0.0$ (almost half filling, $n_{\rm e} \simeq 1.0$), the spin structure factor shows a sharp peak at $\bm{Q}=(\pi, 0, 0)$ for both $G=0$ and $G=0.5$ (not shown here). 
The peak values do not depend on the system size, which suggests the UD order is stable even when allowing other magnetic orders in larger unit cells than in the variational calculations. 
Similar to the results in Sect.~\ref{sec:Variational Calculation}, the moments for the UD order are aligned in the $z$ direction when introducing $G$. 

As shown in Fig.~\ref{Fig:annealing_Kondo_zigzag}(a), the peaks remain at $\bm{Q}=(\pi, 0, 0)$ against a slight decrease of $n_{{\rm e}}$ by a few percent. 
However, the peak for $G=0$ slightly decreases ($\sim 1.5\%$) when increasing the system size from $N=16$ to $24$, while the peak value for $G=0.5$ remains almost unchanged (the change is less than 0.3\%).
These results imply that the UD state survives against a slight doping, and the antisymmetric exchange couplings stabilize the $z$-UD state.

While further decreasing $n_{{\rm e}}$, as shown in Fig.~\ref{Fig:annealing_Kondo_zigzag}(b), the peak position for $G=0$ shifts to a smaller $q_x$. 
This suggests the commensurate UD state is no longer stable and taken over by an incommensurate order with a longer period, as anticipated in the variational arguments in the previous subsection. 
Meanwhile, for $G=0.5$, the $z$-UD state remains stable, as shown in Fig.~\ref{Fig:annealing_Kondo_zigzag}(b). 
Thus, the results of the simulated annealing confirm the variational results in the previous subsection: the $z$-UD ordered state appears as a stable phase at and near half filling in the presence of the antisymmetric exchange couplings. 

\subsubsection{Monte Carlo simulation}
\label{sec:Monte Carlo Simulation}

In this subsection, we investigate the stability of the $z$-UD phase at finite temperatures by Monte Carlo simulation. 
In the Monte Carlo calculations, we adopt the standard method for the spin-charge coupled systems with classical localized moments~\cite{Yunoki_PhysRevLett.80.845}. 
We typically performed 
10,000-90,000 Monte Carlo steps after 10,000 steps for thermalization. 
The statistical errors were estimated by dividing the data into five to ten bins and calculating the standard deviation among the bins. 
The calculations were performed on the $N=4 L \times L \times L$-site lattice with $L=2$ and $4$ under the periodic boundary conditions in all the directions. 
For $L=2$, we introduced a supercell consisting of $N_k = 2^3$ copies of the $N$-site lattice so that the finite-size effect is reduced. 
In order to stabilize the $q_y=q_z=0$ order as in the previous sections, 
we introduce the additional ferromagnetic exchange interaction between localized spins $\mathcal{H}_{{\rm F}} = -J_{{\rm F}} \sum_{\langle i, j \rangle_{yz}} \bm{S}_i \cdot \bm{S}_j$; 
we take $J_{{\rm F}} = 0.1$ and the sum of $\langle i,j \rangle_{yz}$ over the nearest-neighbor sites for the $y$ and $z$ directions. 

\begin{figure}[htb!]
\begin{center}
\includegraphics[width=0.9 \hsize]{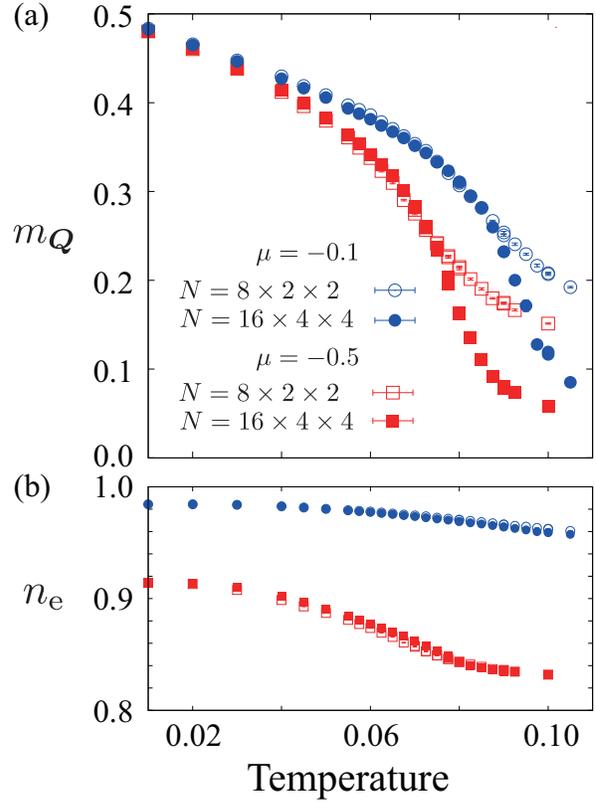} 
\caption{
\label{Fig:zika_Monte}
Monte Carlo results for (a) the order parameter for the $z$-UD order, $m_{\bm{Q}}$, and (b) the electron density $n_{{\rm e}}$ at $\mu=-0.1$ and $\mu = -0.5$. 
The calculations were done at $t_1=1$, $t_2=0.1$, $t_3=0.2$, $t_4=0.2$ $J=4$, and $G=0.5$ for the system sizes $N=8 \times 2 \times 2$ and $16 \times 4 \times 4$.
}
\end{center}
\end{figure}

\begin{figure}[htb!]
\begin{center}
\includegraphics[width=0.9 \hsize]{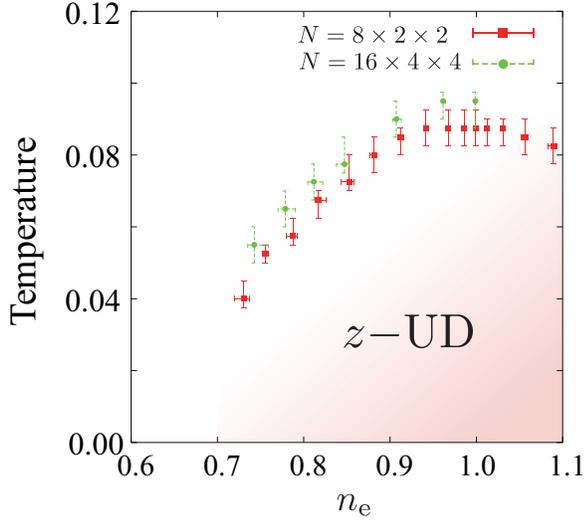} 
\caption{
\label{Fig:souzu_Monte}
Finite-temperature phase diagram of the model in Eq.~(\ref{eq:Ham_zigzag}) as a function of the electron density $n_{{\rm e}}$ obtained by the Monte Carlo simulation. 
The symbols indicate the critical temperatures, which are estimated by the inflection points of the order parameter $m_{\bm{Q}}$ plotted in Fig.~\ref{Fig:zika_Monte}. 
The horizontal error bars indicate the statistical errors of $n_{\rm e}$ at the chemical potential for the critical temperature. 
The results are calculated at $t_1=1$, $t_2=0.1$, $t_3=0.2$, $t_4=0.2$, $J=4$, and $G=0.5$. 
}
\end{center}
\end{figure}

Figure~\ref{Fig:zika_Monte}(a) shows the temperature dependence of the order parameter for the UD order, $m_{\bm{Q}}$. 
Here, $m_{\bm{Q}}$ is defined by $m_{\bm{Q}}=[S_{\bm{Q}}/N]^{1/2}$, where $S_{\bm{Q}}$ is the spin structure factor in Eq.~(\ref{eq:Sq_sec3}) at $\bm{Q} =(\pi,0,0)$. 
The data are calculated for the two different system sizes at $J=4$ and $G=0.5$ for two different values of the chemical potential, $\mu=-0.1$ and $\mu=-0.5$. 
The results show that $m_{\bm{Q}}$ develops rapidly below a particular temperature. 
We confirmed that the magnetic moments are along the $z$ direction by analyzing the spin component of $S_{\bm{Q}}$. 
These indicate a phase transition from the high-temperature 
paramagnetic state to the low-temperature $z$-UD ordered state. 
The rough estimates of the critical temperature $T_c$ can be obtained from the inflection point of $m_{\bm{Q}}(T)$: 
$T_c \simeq 0.09$ at $\mu=-0.1$ ($n_{{\rm e}}\sim 0.96$) and $T_c \simeq 0.075$ at $\mu=-0.5$ ($n_{{\rm e}}\sim 0.85$) [see also Fig.~\ref{Fig:zika_Monte}(b)]. 
With further decreasing temperature, the order parameter $m_{\bm{Q}}$ approaches its saturated value $0.5$ in the ground state. 
This confirms that the multipole ordered state found in the variational calculations remains stable against thermal fluctuations as well as the carrier doping.

By calculating $m_{\bm{Q}}(T)$ while varying $\mu$ in a similar way, we obtain the finite-temperature phase diagram in Fig.~\ref{Fig:souzu_Monte}.
The $z$-UD phase is stable around half filling, with a dome-like shape of $T_c$ with its maximum around $n_{\rm e}=1$. 
Although it is difficult to determine the phase boundary at $n_{{\rm e}} \sim 0.7$ within the limited system sizes in the current calculations, 
the result indicates that the multipole ordered state is robustly stable around half filling, consistent with the results by the variational calculation and simulated annealing.

\section{Summary}
\label{sec:Summary}

In summary, we have investigated the odd-parity multipole ordering that is spontaneously induced by the antisymmetric spin-orbit coupling in the systems with local parity mixing. 
Starting from a general form of the site-dependent antisymmetric hybridization between different parity orbitals, we derived an effective low-energy model with site-dependent antisymmetric exchange couplings. 
This is an extended Kondo lattice model, which is a fundamental model for considering the effect of antisymmetric hybridization in $d$- and $f$-electron systems. 
We have analyzed the model on a quasi-one-dimensional zig-zag lattice, as the minimal lattice structure describing local parity mixing. 
We found that the antisymmetric exchange couplings induce the effective hopping of conduction electrons depending on the configurations of localized spins. 
One of the stable configurations is a N\'eel type antiferromagnetic order with the moments perpendicular to the zig-zag plane. 
The N\'eel order accompanies an odd-parity multipole order composed of magnetic toroidal and quadrupole components. 
This unusual multipole order exhibits a band deformation with a band bottom shift and magnetoelectric response, due to the activated toroidal moment. 
We have investigated the stability of the multipole ordered state by the complementary numerical calculations, i.e., the variational calculation for the ground state, the simulated annealing, and the Monte Carlo simulation at finite temperatures. 
We found that the multipole ordered state is indeed stabilized by the antisymmetric exchange couplings in a wide parameter range at and near half filing. 

Our results will stimulate further studies of odd-parity multipole ordering in systems with local parity mixing. 
Multipole orders similar to that in the present study will be widely observed in the materials to meet the following conditions: (i) local inversion symmetry breaking due to the lattice structure (zig-zag, honeycomb, diamond, \dots), (ii) strong spin-orbit coupling, (iii) hybridization between different parity orbitals, e.g., $s$-$f$, $p$-$d$, and $d$-$f$. 
There are many candidate materials to meet such conditions in $f$-electron systems, as mentioned in the introduction. 
We anticipate a similar situation also in 4$d$- and 5$d$-electron systems where localized levels are expected under some crystal field splitting.
It is desired to systematically study such systems from the viewpoint of odd-parity multipole ordering for further understanding of the exotic magnetism, accompanying peculiar electronic and transport properties. 

\begin{acknowledgements}
SH is supported by Grant-in-Aid for JSPS Fellows. 
This work was supported by Grants-in-Aid for Scientific Research (No.~24340076), the Strategic Programs for Innovative Research (SPIRE), MEXT, and the Computational Materials Science Initiative (CMSI), Japan. 
\end{acknowledgements}

\appendix

\section{Canonical Transformation in the Presence of Antisymmetric Hybridization}
\label{sec:Kondo Lattice Model with ASOC}

In this Appendix, we derive the extended Kondo lattice model in Eq.~(\ref{eq:Ham_zigzag}) 
from the extended periodic Anderson model with the antisymmetric hybridization in Eq.~(\ref{eq:Ham_PAMSO}). 
Following the procedure for deriving the standard Kondo lattice model by the Schrieffer-Wolff transformation~\cite{Schrieffer_PhysRev.149.491}, 
we treat the hybridization $\mathcal{H}_1$ in Eq.~(\ref{eq:Ham_PAMSO2}) as a perturbation to $\mathcal{H}_0$ in Eq.~(\ref{eq:Ham_PAMSO1}), and perform the second-order perturbation expansion by using a canonical transformation.
The canonical transformation is represented by 
\begin{align}
\label{eq:Schrieffer-Wolff}
\tilde{\mathcal{H}} = e^S \mathcal{H} e^{-S}, 
\end{align} 
where $S$ is the anti-Hermitian operator, determined so as to satisfy the following relation:
\begin{align}
\label{eq:repS_zero}
\mathcal{H}_1 + [S, \mathcal{H}_0] =0. 
\end{align}
Here, $[\cdots]$ represents a commutator. 
Expanding Eq.~(\ref{eq:Schrieffer-Wolff}) 
up to the second-order in $\mathcal{H}_1$, we end up with an effective model of Kondo lattice type: 
\begin{align}
\label{eq:Kondo_Ham_SW}
\tilde{\mathcal{H}}_{{\rm ex-KLM}} =  \mathcal{H}_0 + \frac{1}{2} [S, \mathcal{H}_1]. 
\end{align}
Note that $S$ is ${\cal O}(\mathcal{H}_1)$ due to the relation in Eq.~(\ref{eq:repS_zero}). 

Specifically, the operator $S$ is obtained in the form: 
\begin{align}
\label{eq:rep_S}
S &=  
\sum_{l,l',p, \bm{k}, \sigma,\sigma'} 
\left\{ A_{l\sigma \sigma'}(\bm{k}) B_{pl' -\sigma'} (\bm{k}) c^{\dagger}_{l\bm{k}\sigma}f^{}_{pl'\sigma'} -{\rm H.c.} 
\right\}, 
\end{align}
where
\begin{align}
A_l(\bm{k}) &= \left(
\begin{array}{cc}
A_{l\uparrow \uparrow}(\bm{k})
 &  A_{l\uparrow \downarrow}(\bm{k}) \\
A_{l\downarrow \uparrow}(\bm{k})
 & A_{l\downarrow \downarrow}(\bm{k})
\end{array}
\right), \\
&= 
 V_l(\bm{k}) \sigma^0 +  \bm{g}_l^{cf} (\bm{k}) \cdot \bm{\sigma}, \\
B_{pl\sigma}(\bm{k}) &=\left\{ P(\bm{k}) +Q(\bm{k}) n^{f}_{pl\sigma} \right\} e^{-{\rm i} \bm{k}\cdot \bm{R}_p}, 
\\ 
P(\bm{k}) & =\frac{1}{\tilde{\varepsilon}({\bm{k}})-E_0}, \\
Q(\bm{k}) & =\frac{1}{\tilde{\varepsilon}({\bm{k}})-E_0}-\frac{1}{\tilde{\varepsilon}({\bm{k}})-E_0-U}. 
\end{align}
$\tilde{\varepsilon}(\bm{k})= \sum_{l l'} \varepsilon_{ll'}(\bm{k})$ is the energy dispersion of conduction electrons, which is obtained by the Fourier transform of the first term in Eq.~(\ref{eq:Ham_PAMSO}) [see below in Eq.~(\ref{eq:Kondo_Ham_ASOC})], and $\bm{R}_p$ is the position vector for unit cell $p$. 
By substituting $S$ in Eq.~(\ref{eq:rep_S}) into the second term in Eq.~(\ref{eq:Kondo_Ham_SW}), we obtain the effective Hamiltonian, which is represented by
\begin{align}
\label{eq:Kondo_Ham_ASOC}
\mathcal{H}_{{\rm ex-KLM}}  &= \sum_{l,l',\bm{k},\sigma} 
\varepsilon_{ll'} (\bm{k}) c^{\dagger}_{l\bm{k} \sigma} c^{}_{l'\bm{k} \sigma} + \sum_{l,p, \bm{k},\bm{k'}}   
J_{l\bm{k}' \bm{k}} \bm{S}_{pl} \cdot \bm{s}_{l\bm{k}'\bm{k}} \delta_{\bm{R}_p}
\nonumber \\
+\frac{1}{2} \sum_{l,p, \bm{k},\bm{k'}}  
&\left[ \left\{ D^{x}_{l\bm{k}'\bm{k}} 
\left(S_{pl}^{+} n_{l\bm{k}' \bm{k} \downarrow} 
+S_{pl}^{-} n_{l\bm{k}' \bm{k} \uparrow} 
+2 {\rm i} S_{pl}^{z}s^{y}_{l\bm{k}'\bm{k}} 
\right) \right. \right. \nonumber \\ 
&\left. \left. - {\rm i}
 D^{y}_{l{\bm{k}' \bm{k}}}
\left( S_{pl}^{+}n_{l\bm{k}' \bm{k} \downarrow} 
- S_{pl}^{-}n_{l\bm{k}' \bm{k} \uparrow} 
+ 2 S_{pl}^{z}s^x_{l\bm{k}' \bm{k}} 
\right) \right. \right.  \nonumber \\ 
&\left. \left. +
 D^{z}_{l{\bm{k}' \bm{k}}}
\left( S_{pl}^{+}s^-_{\bm{k}' \bm{k}}
-S_{pl}^{-}s^+_{l\bm{k}' \bm{k}}
+S_{pl}^{z}n_{l\bm{k}' \bm{k}}
\right) \right\} \delta_{\bm{R}_p} + {\rm H.c.}\right] \nonumber \\ 
+ \frac{1}{2}\sum_{l,p, \bm{k},\bm{k'}} 
&\left\{
G^{xx}_{l{\bm{k}'\bm{k}}} 
\left( S_{pl}^{+}s^+_{l\bm{k}' \bm{k}}
+S_{pl}^{-}s^-_{l\bm{k}' \bm{k}}
-2S_{pl}^{z}s^z_{l\bm{k}' \bm{k}}
\right) \right.  \nonumber \\ 
&\left. -
 G^{yy}_{l{\bm{k}'\bm{k}}} 
\left(S_{pl}^{+}s^+_{l\bm{k}' \bm{k}}
+S_{pl}^{-}s^+_{l\bm{k}' \bm{k}}
+2S_{pl}^{z}s^z_{l\bm{k}' \bm{k}}
\right) \right. \nonumber \\
&\left.  -  G^{zz}_{l{\bm{k}'\bm{k}}} 
\left( S_{pl}^{+}s^-_{l\bm{k}' \bm{k}}
+S_{pl}^{-}s^+_{l\bm{k}' \bm{k}}
-2S_{pl}^{z}s^z_{l\bm{k}' \bm{k}}
\right) \right\} \delta_{\bm{R}_i}  \nonumber \\
+\frac{1}{2} \sum_{l,p, \bm{k},\bm{k'}} 
&\left[\left\{
- G^{xz}_{l\bm{k}'\bm{k}} 
\left(S_{pl}^{+}n_{l\bm{k}' \bm{k} \downarrow} 
-S_{pl}^{-}n_{l\bm{k}' \bm{k} \uparrow} 
-2S_{pl}^{z}s^x_{l\bm{k}' \bm{k}}
\right)  \right. \right. \nonumber \\ 
&\left. \left. -{\rm i}
  G^{xy}_{l\bm{k}' \bm{k}} 
\left( S_{pl}^{+}s^+_{l\bm{k}' \bm{k}}
- S_{pl}^{-}s^-_{l\bm{k}' \bm{k}}
+ S_{pl}^{z}n_{l\bm{k}' \bm{k}}
\right)  \right. \right. \nonumber \\ 
&\left. \left. +{\rm i}
 G^{yz}_{l\bm{k}' \bm{k}} 
\left( S_{pl}^{+}n_{l\bm{k}' \bm{k} \downarrow}
+ S_{pl}^{-}n_{l\bm{k}' \bm{k} \uparrow}
- 2 {\rm i} S_{pl}^{z}s^y_{l\bm{k}' \bm{k}}
\right) \right\}\delta_{\bm{R}_p}+ {\rm H.c.}\right], \nonumber \\ 
\end{align}
where
$n_{l\bm{k}'\bm{k}}=\sum_{\sigma}n_{l\bm{k}'\bm{k}\sigma}$ with $n_{l\bm{k}'\bm{k}\sigma}=c^{\dagger}_{l\bm{k}'\sigma}c^{}_{l\bm{k}\sigma}$, 
$s^{x}_{l\bm{k}'\bm{k}}=(s^{+}_{l\bm{k}'\bm{k}}+s^{-}_{l\bm{k}'\bm{k}})/2$, and $s^{y}_{l\bm{k}'\bm{k}}=
(s^{+}_{l\bm{k}'\bm{k}}-s^{-}_{l\bm{k}'\bm{k}})/2{\rm i}$. 
The coefficients for the exchange couplings are given by 
\begin{align}
\label{eq:J_expression}
J_{l\bm{k}' \bm{k}} &= V_l(\bm{k}') V_l^{*}(\bm{k}) Q_{{\bm{k},\bm{k}'}}, \\
\label{eq:D_expression}
\bm{D}_{l\bm{k}' \bm{k}} &= V_l(\bm{k}') [\bm{g}_l^{cf}(\bm{k})]^* Q_{{\bm{k},\bm{k}'}}, \\
\label{eq:G_expression}
G^{\mu \nu}_{l\bm{k}' \bm{k}} &= [g_l^{cf,\mu}(\bm{k}')] [g_l^{cf,\nu}(\bm{k})]^* Q_{{\bm{k},\bm{k}'}},
\end{align}
where $Q_{{\bm{k},\bm{k}'}} = Q(\bm{k}) + Q(\bm{k}')$ and $\mu, \nu = x, y, z$. 
In the derivation, we used the condition $n^f_i=1$, assuming the $f$ electrons are well localized in a singly-occupied state at each site, as the standard Kondo lattice model. 
In the same spirit, we neglect pair hopping terms proportional to $c^{\dagger}_{i \sigma}c^{\dagger}_{j \sigma}f^{}_{i\sigma}f^{}_{j \sigma}$ and one-body terms for $f$ electrons proportional to $f^{\dagger}_{i\sigma}f^{}_{j \sigma}$~\cite{comment_one_body}. 

Equation~(\ref{eq:Kondo_Ham_ASOC}) provides a general form of the extended Kondo lattice model applicable to any lattice structure. 
Finally, we simplify the $\bm{k}$ dependence in Eq.~(\ref{eq:Kondo_Ham_ASOC}), bearing a 
quasi-one-dimensional system composed of
zig-zag chains [see Fig.~\ref{Fig:zigzag_chain}(b)] in mind. 
Namely, we drop all the $\bm{k}$ dependences except for the most interesting one, $\sin k_x$ in $\bm{g}_l^{cf}(\bm{k})$ in Eq.~(\ref{eq:gsimple}), as 
\begin{align}
\label{eq:J_simple}
J_{l\bm{k}' \bm{k}} &\rightarrow J, \\
\label{eq:D_simple}
D^{\mu}_{l\bm{k}' \bm{k}} &\rightarrow D_l\, \delta_{\mu z} \sin k_x, \\
\label{eq:G_simple}
G^{\mu \nu}_{l\bm{k}' \bm{k}} &\rightarrow G\, \delta_{\mu z}\, \delta_{\nu z} \sin k_x \sin k'_x, 
\end{align}
where $\delta_{\mu \nu}$ is the Kronecker delta.
Here, $J$ and $G$ are generally positive from Eqs.~(\ref{eq:J_expression}) and (\ref{eq:G_expression}). 
Note that $D_l$ depends on sublattice index, which comes from 
$\bm{g}^{cf}_{l}(\bm{k})\propto (-1)^{l}$ for the present zig-zag lattice structure.
We also approximate the magnitude of $D_l$ by $\sqrt{JG}$ from Eqs.~(\ref{eq:J_expression}), (\ref{eq:D_expression}), and (\ref{eq:G_expression}). 
By substituting Eqs.~(\ref{eq:J_simple}), (\ref{eq:D_simple}), and (\ref{eq:G_simple}) into Eq.~(\ref{eq:Kondo_Ham_ASOC}), we obtain the extended Kondo lattice Hamiltonian for the zig-zag lattice structure as in Eq.~(\ref{eq:Ham_zigzag}).

\bibliographystyle{JPSJ}
\bibliography{ref}

\begin{thebibliography}{10}
\bibitem{comment_wave_number}
In this and next subsections, we redefine the wave vector; we suppose that the
  chain is straight in the $x$ direction, and take the primitive translation
  vector as that for the nearest-neighbor sites (connecting the neighboring A
  and B sublattices).
\bibitem{comment_one_body}
Except for the $c^{\dagger}_{i \sigma}c^{\dagger}_{i
  \sigma}f^{}_{i\sigma}f^{}_{i \sigma}$ and $f^{\dagger}_{i\sigma}f^{}_{i
  \sigma}$ terms, one-body terms for $c$ electrons proportional to
  $c^{\dagger}_{i\sigma}c^{}_{i \sigma}$ appear within the second-order
  perturbation. We here omit these terms because they are canceled out by the
  constant terms appearing from the exchange coupling terms in
  Eq.~(\ref{eq:Kondo_Ham_ASOC}) in the strong $U$ limit.
\end{thebibliography}

\end{document}